\documentclass[a4paper,oneside,english]{aa}
\usepackage{times}
\usepackage[T1]{fontenc}
\usepackage{amsmath}
\usepackage{graphicx}
\usepackage{amssymb}
\usepackage{xspace}

\newcommand{\ascc}   {\mbox{ASCC-2.5}\xspace}

\newcommand{\mc}[3]{\multicolumn{#1}{#2}{#3}}

\usepackage{babel}
\makeatother
\begin{document}

\title{Towards absolute scales of radii and masses of open clusters}

\author{A.E.~Piskunov \inst{1,2,3}\and
        E.~Schilbach \inst{1} \and
        N.V.~Kharchenko \inst{1,3,4}\and
        S.~R\"{o}ser \inst{1}\and
        R.-D.~Scholz \inst{3} }

\offprints{R.-D.~Scholz}

\institute{Astronomisches Rechen-Institut, M\"{o}nchhofstra\ss{}e 12-14,
D--69120 Heidelberg, Germany\\
email: apiskunov@ari.uni-heidelberg.de, elena@ari.uni-heidelberg.de,
nkhar@ari.uni-heidelberg.de, roeser@ari.uni-heidelberg.de
\and
Institute of Astronomy of the Russian Acad. Sci., 48 Pyatnitskaya
Str., 109017 Moscow, Russia\\
email: piskunov@inasan.rssi.ru
\and
Astrophysikalisches Institut Potsdam, An der Sternwarte 16, D--14482
Potsdam, Germany\\
email: apiskunov@aip.de, nkharchenko@aip.de, rdscholz@aip.de
\and
Main Astronomical Observatory, 27 Academica Zabolotnogo Str., 03680
Kiev, Ukraine\\
email: nkhar@mao.kiev.ua
}

\date{Received 10 January 2007; accepted ...}

\abstract{}{
In this paper we derive tidal radii and masses of open clusters in the nearest
kiloparsecs around the Sun.}
{For each cluster, the mass is estimated from tidal radii determined from a
fitting of three-parametric King's profiles to the observed integrated density
distribution. Different samples of members are investigated.}
{For 236 open
clusters, all contained in the catalogue \ascc, we obtain
core and tidal radii, as well as tidal masses.
The distributions of the core and tidal radii peak at about 1.5~pc and
7 - 10~pc, respectively. A typical relative error of the core radius lies
between 15\% and 50\%, whereas, for the majority of clusters, the tidal
radius was determined with a relative accuracy better than 20\%. Most
of the clusters have tidal masses between 50 and 1000 $m_\odot$, and
for about half of the clusters, the masses were obtained with a relative error
better than 50\%.}{}


\maketitle

\section{Introduction}

As a minimum characteristics to describe a stellar cluster one needs to specify
the position of a cluster centre and its apparent (angular) size. Both
parameters serve
as a means of identifying a cluster on the sky. Also, these parameters are the
most
common, present in numerous catalogues of clusters, and presently available for
about 1700 galactic open clusters (see e.g. Dias et al.~\cite{daml02}). In the
majority of
cases,
however, they are derived from visual inspection of the area of a cluster on the
sky.
So, the parameters may be strongly biased due to the size of the detector field,
and/or
by contamination of field stars, and, hence,
present a
lower limit of the real size of a cluster. As shown by Kharchenko et
al.~(\cite{clucat}) these  data from literature are normally smaller by a factor
of two
with respect to cluster radii drawn from the analysis of a uniform
membership based on photometric and spatio-kinematic constraints. These latter
data are in turn subject to various biases (see Schilbach et
al.~\cite{clusim}) and need to be reduced to a uniform system in order to allow
physical insight into structural properties of the population  of galactic open
clusters.

Besides the morphological description of a cluster itself, the
structural parameters carry important information on its basic physical
properties like mass, and on the surrounding galactic tidal field (von
Hoerner~\cite{hoern57}). King~(\cite{king62}) has proposed an empirical set of
cluster parameters
as a quantifier of the
structure of spherical systems, and
has shown later (King~\cite{king66}) that they correspond to theoretical density
profiles of quasi-equilibrium configurations and suit well to stellar clusters
with
masses spanning from those of open clusters to globular ones.
King~(\cite{king62})
introduced three spatial parameters $r_c,\,r_t$ and $k$, hereafter referred to
as
King's parameters ($r_c$ is the so-called core radius,  $r_t$ the tidal radius,
and $k$
is a profile normalization factor), fully describing the distribution of the
projected density in a cluster. Since that time, this parameter set is widely
used to quantitatively describe the
density laws of globular clusters. Here we  mention a number of studies, where
King's parameters were determined both for galactic globular clusters (e.g.
Peterson and King~\cite{peter75}, Trager et al.~\cite{trag96}, Lehmann \&
Scholz~\cite{lehm97}), and for extragalactic ones (Kontizas~\cite{kont84}, Hill
\&
Zaritski~\cite{hill06}) in the SMC, or LMC (Elson et al.~\cite{els87}).

The application of King's parameters might be useful to open
clusters as well, especially from the point of view of establishing a uniform
scale of
structural parameters and providing independent estimates of cluster
masses.
However, the literature on the determination of King's parameters of open
clusters is
much poorer than that of globulars. We mention here the following studies  based
on
three-parameter fits of selected clusters: King~(\cite{king62}),
Leonard~(\cite{leo88}), Raboud \& Mermilliod~(\cite{ramer98a},
\cite{ramer98b}). Among other complicating reasons such as an insufficient
stellar
population, and heavy and irregular fore/background, making the study of
King's parameters in open clusters difficult, we emphasise the difficulty of
acquiring
data in wide-field areas around a cluster. The latter is due to the much larger
size of a typical open cluster compared to the fields of view of
detectors currently used in studies of individual clusters.
Recently, when a number of all-sky catalogues has become available,
studies exploring the unlimited neighbourhood of clusters  have been published
(Adams et al.~\cite{adea01}, \cite{adea02}, Bica et al.~\cite{bonbs},
\cite{bonb},~\cite{bibob}). Froebrich
et al.~(\cite{froeb07}) have been searching the 2MASS survey for new clusters,
and provided spatial parameters for all newly identified cluster candidates
derived from the fitting of surface density patterns with King profiles.

Mass is one of the fundamental parameters of star clusters. There
are several independent methods to estimate cluster masses.
Each of them has its advantages or disadvantages with respect to the
other ones. But so far there is no method, which can be regarded as absolutely
satisfactory.
The most simple and straightforward way is to count cluster members and to sum
up
their masses. Since there is no cluster with a complete census of members, one
always
observes only a subset of cluster stars, truncated by the limiting magnitude and
by the
limited area covered by a study, and, therefore, masses from star counts should
be regarded
as lower estimates of real mass of a cluster. The extrapolation of the
mass spectrum to an unseen lower limit of stellar masses along with some
template of the IMF frequently applied in such studies, leads to
unjustified and unpredictable modifications of the observed mass and should be
avoided. The farther away the cluster is, the larger is the uncounted fraction
of faint members, often residing in the cluster periphery.
In fact, this method could be applied with reasonable
degree of safety to the nearest clusters observed with deep, wide-area surveys,
and so
providing secure and complete membership. Due to its simplicity the method is
currently widely accepted, and possibly it is the only technique which is
applied
to relatively large samples of open clusters (see Danilov \&
Seleznev~\cite{danil},
Tadross et al.~\cite{tad02}, and Lamers et al.~\cite{lamea}).

The second method is the classical one, namely
the application of the virial theorem. It gives the mass of a cluster
from an estimate of the
stellar velocity dispersion, and average interstellar distances. It
does not require the observation and membership determination of all cluster
stars.
The application of the method is,
however, limited to sufficiently massive stellar systems (globulars and
dwarf spheroidals)
with dispersions of internal motions large enough to be measurable. For open
clusters with typcal dispersions of the order of or less than 1 km\,s$^{-1}$
present-day
accuracies of both proper motions and radial velocities are fairly rough
and are marginally available for a few selected clusters only. In spite of
this, several attempts have been undertaken for clusters with the most accurate
proper motions
(Jones~\cite{jones70}, \cite{jones71}, McNamara \& Sanders~\cite{mcn77},
\cite{mcn83}, McNamara \& Sekiguchi~\cite{mcn86}, Girard et al.~\cite{gir89},
Leonard \& Merritt~\cite{leo89}), or for clusters with mass determination from
radial velocities (Eigenbrod et al.~\cite{eigen}).

The third method uses the interpretation of the tidal interaction of a cluster
with
the parent galaxy, and requires the knowledge of the tidal radius of a cluster.
Considering globular clusters which, in general, have elliptical orbits,
King~(\cite{king62}) differentiates between the tidal and the limiting
radius of a cluster. For open
clusters revolving at approximately circular orbits one can expect the observed
tidal radius being approximately equal to the limiting one. Though a probable
deviation of the cluster shape from sphericity may have some impact onto the
computed
cluster mass.
Nevertheless, this method gives a mass estimate of a cluster
(Raboud \& Mermilliod~\cite{ramer98a}, \cite{ramer98b}) which is independent
from the
results of the two methods mentioned above.
Due to the cubic dependence on $r_{t}$, masses drawn from tidal radii are
strongly influenced
by the uncertainties of $r_{t}$, however.
Taking these circumstances into account, one usually reverses the relation and
calculates tidal radii from counted masses.

For our studies of open clusters we use the All-Sky Compiled Catalogue
of 2.5 million stars\footnote{ftp://cdsarc.u-strasbg.fr/pub/cats/I/280A} (\ascc,
Kharchenko~\cite{kha01}), including absolute proper motions in
the Hip\-par\-cos system, $B$, $V$ magnitudes in the Johnson photometric
system, and supplemented with spectral types and radial velocities
if available. The \ascc is complete down to about
$V=11.5$~mag. Based on the \ascc we were able to construct reliable
combined kinematic-photometric membership probabilities of bright stars
($V\lesssim 12$) for 520 open clusters (Kharchenko et al.~\cite{starcat},
Paper~I), to compute a uniform set of astrophysical parameters of clusters,
(Kharchenko et al.~\cite{clucat}, Paper~II), as well as to identify 130
new clusters (Kharchenko et al.~\cite{newclu}, Paper~III) in \ascc. Currently,
we have a
sample of 650 open clusters, which is complete within a distance of about 1 kpc
from the Sun.
This sample was used to study the population of open clusters in the
local Galactic disk by jointly analysing the spatial and kinematic
distributions of clusters (Piskunov et al.~\cite{clupop}, Paper
IV), for an analysis of different biases affecting the apparent size of open
clusters, and the segregation of stars of different mass in open clusters
(Schilbach et al.~\cite{clusim}, Paper~V).

In this paper we determine King's parameters and tidal masses
for a large fraction of open
clusters from our sample to get an independent basis for the
construction of a uniform and objective scale of spatial parameters and masses.
In Sec.~\ref{sec:data} we briefly discuss our input data,
Sec.~\ref{sec:detrm} contains the description of the pipeline we apply for the
determination
of King's parameters, in Sec.~\ref{sec:res} we construct and discuss
our sample, in Sec.~\ref{sec:compar} we compare our results with published data
on King's parameters and with independent estimates of cluster masses. In
Sec.~\ref{sec:concl} we summarize the results.

\begin{figure}
\includegraphics[%
  width=1.0\columnwidth%
  ]{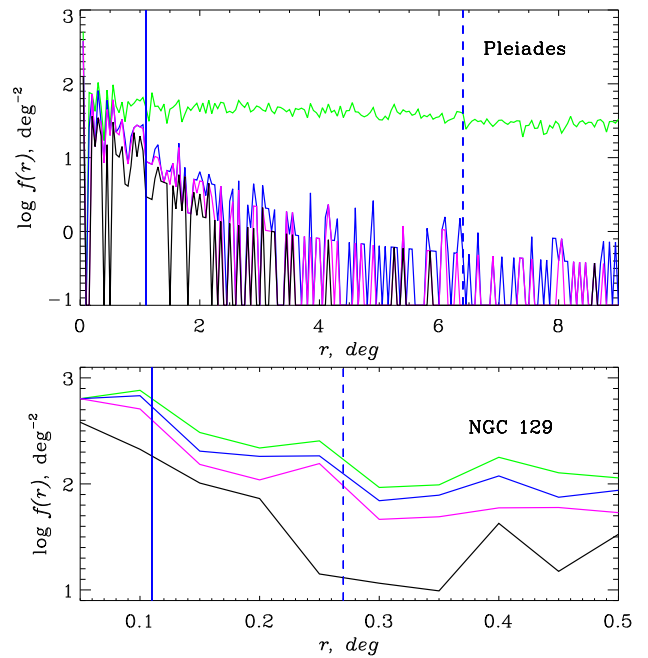}

\caption{Apparent density profiles of two open clusters used for the
determination of tidal radii. The Pleiades are shown in the top
panel, NGC~129 in the bottom panel. The different colors indicate
different samples of stars considered: black for $1\sigma$-members
(group (a), see text), magenta for $2\sigma$-members (group (b)),
blue for $3\sigma$-members (group (c)), and green for
``$4\sigma$-members'' (group (d)).  Vertical lines show the radii
$r_1$ (solid) and $r_2$ (dashed).}\label{fig:obspro}
\end{figure}

\section{Data}\label{sec:data}

In this study, we make use of our results on the cluster membership
(Paper~I), and on
the parameter determination (Paper~II and Paper~III, respectively)
for 650 open clusters.  Together with other basic parameters like
the position of the cluster centre, age, distance, and angular size
(the apparent radii of the core $r_1$ and the corona $r_2$), stellar
density
profiles in the wider neighbourhood of each cluster are available from
our data.  According to the membership probability $P_{ap}$\footnote{$P_{ap}$
is a newly defined combined membership probability computed for each star
in a cluster area by taking into account only the kinematic and photometric
criteria, without spatial selection criteria,
$P_{ap}=\mathrm{min}\{P_\mu,P_{ph}\}$. Note that this definition of the
combined membership probability differs from that of $P_c$ used in Paper~I.},
density profiles have been constructed for four
groups of stars in each cluster:
(a) - the most probable members ($P_{ap}>61$\%), so called $1\sigma$-members,
(b) - possible members ($P_{ap}>14$\%), or $2\sigma$-members,
(c) - stars with $P_{ap}>1$\%, or $3\sigma$-members, and finally,
(d) - all stars in the cluster area which, for convenience, we call
 ``$4\sigma$-members''.

For the construction of the density profiles, we count stars in
concentric rings around the cluster centre up to $5\,r_2$,
where $r_2$ is the apparent radius of the corona.
Due to the relatively
bright magnitude limit of the ASCC-2.5, the number
of cluster members available for the profile construction is rather low
(on average, about 45 $2\sigma$-members per cluster). In general, the
number of identified cluster members decreases with increasing distance
modulus of a cluster. In order to have a statistically relevant number
of stars per concentric ring, one has to chose steps of larger
linear size for remote clusters. Therefore, we count stars in concentric
rings of equal angular width (0$\fdg$05).
This homogeneous approach makes the linear
profile spacing automatically larger for more distant clusters. 
Since in the following analysis we intend to derive 3 unknown
   King's parameters from the profile fitting, we need at least four bins
   in the observed profile, and therefore, we select only clusters
   with $r_2\geqslant 0\fdg2$ and with more than 10 stars above the
   background level within $r_2$.
Under these constraints, the sample contains 290
clusters, though, not for all of them, an acceptable solution has been
obtained (see Section \ref{sec:detrm}).

In order to give the reader an idea on the quality of the input data for
the determination of tidal radii, we show two
examples of apparent density profiles in Fig.\ref{fig:obspro}:
the Pleiades, one of the ``best-quality'' clusters in our sample (top
panel), and NGC~129, a remote cluster, and one of the ``low-quality''
clusters (bottom panel).  The corresponding data on the density profiles
for all 650 clusters can be found in the Open Cluster Diagrams Atlas
(OCDA) available from the
CDS\footnote{ftp://cdsarc.u-strasbg.fr/pub/cats/J/A+A/438/1163/atlas
and ftp://cdsarc.u-strasbg.fr/pub/cats/J/A+A/440/403/atlas}.

\section{Determination of tidal radii and cluster masses}\label{sec:detrm}

\subsection{Fitting cluster profiles}

The method we apply is based on the well-known empirical model of
King~(\cite{king62}),
describing the observed projected density profiles $f(r)$ in globular
clusters
with three parameters $r_{c}$, $r_{t}$, and $k$:
\begin{equation}
f(r)=k\left\{
\left[1+(r/r_{c})^{2}\right]^{-1/2}-\left[1+(r_{t}/r_{c})^{2}\right]^{-1/2}
\right\}^{2}.                               \label{eq:kingd}
\end{equation}
According to King's definition, $r_{c}$ is the core radius, $r_{t}$ is the
tidal radius (approximately equal to the limiting radius in case of an open
cluster), and $k$ is a normalization factor which is related to the
central density of the cluster. This approach has also been successfully applied
for the determination of the King parameters in several nicely populated open
clusters (see e.g., Raboud \& Mermilliod~\cite{ramer98a},
\cite{ramer98b}). Nevertheless, the direct way of fitting the observed
density distribution by the model (eq.~(\ref{eq:kingd})) does not work
for the majority of clusters of our sample. The main reason is the relatively
bright magnitude limit of the ASCC-2.5 and, consequently, the low number of
cluster members that causes uncertainties in the observed density profiles.
In order to weaken  the influence of poor statistics, we use the integrated form
of King's formula:
\begin{eqnarray}
n(r) & = & \pi\, r_{c}^{2}k\,\left\{
\ln[1+(r/r_{c})^{2}]-4\frac{\left[1+(r/r_{c})^{2}\right]^{1/2}-1}{\left[1+(r_{t}
/r_{c})^{2}\right]^{1/2}}+\right.        \nonumber\\
 &  & \left.+\frac{(r/r_{c})^{2}}{1+(r_{t}/r_{c})^{2}}\right\}.\label{eq:kingi}
\end{eqnarray}
where $n(r)$ is the number of stars within a circle of  radius $r$.
Since the spatial boundaries of observed open clusters are not clearly defined,
and
the proportion of field stars projected on the cluster area is relatively high,
we must take special care of choosing the integration limits and the
background level if
using eq.~(\ref{eq:kingi}).

Contrary to globular clusters, where the statistical errors of empirical
density profiles are negligibly low, and the data fix a model safely
in internal regions of the cluster area, in open clusters a fit based on the
inner area
is less reliable and can lead to a significant bias in the resulting
tidal radius.
Therefore, we must consider the behaviour of the density
profile
in exterior regions of a cluster and even outside the cluster
limits which, a priori, we do not know. On the other hand, as one can see
from eq.~(\ref{eq:kingd}), the value of $f(r)$ increases at
$r>r_t$ and tends to a finite limit $f(r\rightarrow\infty)$, whereas $n(r)$
goes to infinity for $r\rightarrow\infty$ in eq.~(\ref{eq:kingi}). This
contradicts the physical meaning of $n(r)$ since the number of cluster
members should be finite, independently of how far one expands the counts.
Therefore, for a physically correct application of eq.~(\ref{eq:kingi})
one should complement it by a boundary condition
$n(r\geqslant r_t) = N$ where $N$ is the number of cluster stars.
Again, $r_t$ and $N$ are unknown.

\begin{figure}
\includegraphics[%
  clip,
  width=1.0\columnwidth,
  keepaspectratio]{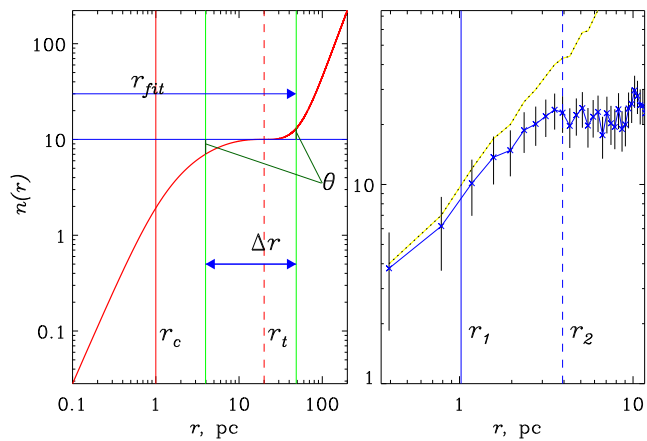}

\caption{The general scheme of fitting cluster density profiles.
Left panel: a theoretical, integrated King's profile for
$r_c=1$ pc, $r_t=20$ pc with the fitting range $r_{fit}$
and the tolerance $\theta$ as
indicated.
Right panel: the density profiles of the open cluster Trumpler~3 from
the $2\sigma$-sample. The measured profile is shown by the short-dashed
line. The empirical profile corrected for the background (see text) is
indicated by crosses, the bars are Poisson errors.
The vertical (blue) solid and dashed lines mark the empirical parameters $r_1$
and $r_2$,
respectively.}
\label{fig:exmpl}
\end{figure}

In order to overcome the problem, we tried to find a range $\Delta r$ where
$n(r)$ is practically flat (see for illustration Fig.~\ref{fig:exmpl}, left
panel). This range includes $r_t$, and
its length depends on the concentration $c=\log r_t/r_c$.
The range $\Delta r$ degenerates to a point $r=r_t$ at $c=0$, and increases
with increasing $c$.

For each cluster, a given membership sample may include a number of field
stars having, by  chance, the proper motions and photometry which are
compatible with those of real cluster members.
Therefore, before we can try to localize $\Delta r$ in the empirical integrated
density profiles, the residual background contamination has to be removed.
If not, the profiles would increase
steadily with increasing $r$. This is especially important for $3\sigma$-
and ``$4\sigma$''-samples which are strongly contaminated by field stars,
although, it can be essential for $1\sigma$ and $2\sigma$-samples, too.

The background correction was done in a uniform way for all clusters
and membership samples. Assuming that
the majority of stars at $r>r_2$ are not cluster members, we took
the average density of a given membership group in a ring $r_2 < r <2\,r_2$
to be the initial background level in the internal cluster area $(r<r_2)$.
Outside $r_2$ $(r_2 < r<5 \, r_2)$, the initial background was first set
to the observed stellar density in each radial bin. The final, smooth
background profile over the complete area $0<r<5\, r_2$ was then recomputed
as a running average of the initial values
with a filter size of 0$\fdg$55 and a step of 0$\fdg$05.
Finally, the background profile was subtracted from the original density
distribution.
An example of an observed profile before and after
background correction is shown in Fig.~\ref{fig:exmpl}, right panel. Except for
a few cases of poor and extended clusters projected on heavy and variable
background, we obtained reasonable results.

The general scheme of the profile fitting is shown in Fig.~\ref{fig:exmpl}(left)
and is explained in the following:

\textbf{Step 1:} we start with the determination of an
initial value $\tilde{r_2}$ of the fit radius $r_{fit}$. Assuming  the empirical
cluster sizes $r_1$ and $r_2$ to be $r_c$ and $r_t$ respectively, we compute
$\tilde{r_2}$ from King's profile of the concentration class $\tilde{c}=\log
r_2/r_1$
as the distance from the cluster centre where the profile does not differ
from $n(r_2)$ by more than an assumed tolerance $\theta$.
The tolerance is chosen
depending on the observation quality of a given cluster, and is defined as
the average of the Poisson errors of the profile outside $r_2$:
$\theta=\langle\sigma_r(r > r_2 )\rangle,\,\,\sigma_r=\sqrt{n(r)}$.

\textbf{Step 2:} now we are able to apply a nonlinear fitting routine
based on the Levenberg-Marquardt
optimization method (Press et al.,~\cite{numrec}) to
eq.~(\ref{eq:kingi})
for $r = 0...\tilde{r_2}$.
As initial-guess parameters we chose $r^0_c=r_1$ and $r^0_t=r_2$, whereas
$k^0$ is obtained from the solution of eq.~(\ref{eq:kingi}) at $r=r_2$.
Empty bins in the differential density distribution were omitted
from the integrated profile fitting.
As a result of successive iterations, we get King's parameters together
with their $rms$ errors and a $\chi^2$-value. The iterations are stopped when
two
successive $\chi^2$-values do not differ by more than $10^{-3}$, and the
solution is accepted when the number of iterations is less than 100.
Then we compute the $\chi^2$-probability function $Q(\chi^2,\nu)$
which can be used as a measure of the goodness of fit. For a
given degree of freedom $\nu$, $Q(\chi^2,\nu)$ is the
probability that the difference between the observations and the fitted
function can be considered random, and their sum of the squares
is allowed to be greater than $\chi^2$. According to Press et
al.~(\cite{numrec}),
the fit can be accepted when $Q(\chi^2,\nu)>0.1$.

Although a choice of $r_{fit} = \tilde{r_2}$ provides a fitting of a
reasonable portion of King's profile to the observations, we
consider this fit range as the lowest limit of $r_{fit}$.

\textbf{Step 3:}
In order to check whether a better convergence can be achieved,
we run the fitting routine (i.e. \textbf{Step 2})
for different $r_{fit}$ ranging from $\tilde{r_2}$ to $3\,r_2$. If
several acceptable solutions (i.e. $Q(\chi^2,\nu)>0.1$) were obtained,
we selected that one which yielded the smallest $rms$ errors in $r_c$ and
$r_t$.

The complete pipeline including the background elimination and the
profile fitting has been applied to all four membership groups.
Although the initial guess  $\tilde{r_2}$ of the fit radius is usually
$\approx 1.5\,r_2$, the final fit radius $r_{fit}$ turns out to be
about $2\,r_2$, and
the best $r_t$ ranges from 1\,$r_2$ to 2\,$r_2$.

However, we must keep in mind that eqs.~(\ref{eq:kingd}), (\ref{eq:kingi})
   assume spherical symmetry in the spatial distribution of cluster stars, 
   whereas a real open cluster is expected to have an elongated form with  
   the major axis directed towards the
   Galactic centre (Wielen \cite{wiel74}). Thus, depending on the orientation
   of the line of sight, an observer measures a projection rather than the
   real size of a cluster, and the values of $r_t$ derived via
   eqs.~(\ref{eq:kingd}), (\ref{eq:kingi}) give, in general, lower limits
   of the tidal radii.

\begin{table}
\caption{Normalized values of King's parameters computed with four
different membership groups}\label{tab:solqua}
\begin{tabular}{ccccc}
\hline
                    & $1\sigma$   & $2\sigma$   &$3\sigma$    & $4\sigma$   \\
\hline
$\bar{r}_c$         &$0.97\pm0.03$&$1.00\pm0.02$&$1.01\pm0.02$&$1.01\pm0.03$\\
$\bar{r}_t$         &$1.05\pm0.01$&$1.00\pm0.01$&$0.98\pm0.01$&$0.97\pm0.01$\\
$\bar{k}$           &$0.64\pm0.02$&$1.02\pm0.03$&$1.11\pm0.02$&$1.24\pm0.03$\\
$\bar{n}_2$         &$0.65\pm0.01$&$1.04\pm0.01$&$1.13\pm0.01$&$1.18\pm0.02$\\
\hline
$\bar{\delta}_{r_c}$
&$1.22\pm0.02$&$0.96\pm0.02$&$0.91\pm0.02$&$0.91\pm0.02 $\\
$\bar{\delta}_{r_t}$
&$1.32\pm0.04$&$0.99\pm0.03$&$0.90\pm0.03$&$0.80\pm0.03$\\
$\bar{\delta}_{k}$
&$1.22\pm0.04$&$0.94\pm0.03$&$0.90\pm0.02$&$0.94\pm0.03$\\
\hline
$\overline{Q}$
&$1.17\pm0.04$&$1.03\pm0.02$&$1.00\pm0.02$&$0.80\pm0.03$\\
\hline
\end{tabular}
\end{table}


\subsection{Cluster masses}\label{sec:methmass}

According to King~(\cite{king62}), the mass $M_{c}$ of a cluster at the
galactocentric distance $R_G$ follows the relation
\begin{equation}
M_{c}=\frac{4\, A\,(A-B)\, r_{t}^{3}}{G},                \label{eq:mcl}
\end{equation}
where $A,B$ are Oort's constants valid for $R_G$,
and $G$ is the gravitational constant (see Standish~\cite{stan95}).
Since the bulk of our clusters is located within 2~kpc from the Sun, the linear
approximation to the velocity field of the disk seems to be reasonable. Thus,
Oort's constants could be easily expressed by their local values
$A_0=14.5\pm0.8$ km/s/kpc, $B_0=-13.0\pm1.1$ km/s/kpc derived in Paper~IV for
the open cluster
subsystem
\[
\begin{array}{lcl}
A&=&A_0 -A_0\,\delta R_G \\
A-B&=&A_0-B_0-2\,A_0\,\delta R_G\\
\end{array}
\]
where $\delta R_G=(R_G-R_{G,0})/R_{G,0}$, and $R_{G,0}=8.5$ kpc.

\begin{table*}
\caption{Sample table of King parameters and tidal masses for 236 open
clusters. The full table is available in machine readable form only.
See text for further explanations.}\label{tab:results}

\begin{tabular}{@{\extracolsep{-6pt}}rlrrrcrccccrrrrrrrrr}

\hline\hline
 COCD &
 Cluster&
 \mc{1}{c}{$l$}&
 \mc{1}{c}{$b$}&
 \mc{1}{c}{$d$}&
 $E_{B-V}$&
 \mc{1}{c}{$\log t$}&
 $r_1$&
 $r_2$&
 $n_s$&
 $i_s$&
 $n_2$&
 $r_c$&
 $\varepsilon_{r_c}$&
 \mc{1}{c}{$r_t$}&
 $\varepsilon_{r_t}$&
 $k$&
 $\varepsilon_{k}$&
 $\log m        $ &
 $\varepsilon_{\log m        }$\\
    &
    &
 \mc{2}{c}{deg} &
 \mc{1}{c}{pc} &
 mag&
 yrs&
 \mc{2}{c}{deg}&
   &
   &
   &
 \mc{4}{c}{pc}&
   &
   &
 \mc{2}{c}{$m_\odot$}\\
 \hline
  3&  Blanco 1          &  14.17& $-$79.02&   269&  0.01& 8.32&  0.70&  2.90& 4&
2&  53&   1.5&   0.2&  20.0&   3.4&   3.2&   0.4&   3.475 & 0.224   \\
  4&  Alessi 20         & 117.64&  $-$3.69&   450&  0.22& 8.22&  0.12&  0.30& 2&
2&  14&   0.5&   0.3&   3.4&   1.3&  10.3&   5.7&   1.152 & 0.504   \\
  8&  NGC 129           & 120.27&  $-$2.54&  1625&  0.55& 7.87&  0.11&  0.27& 4&
1&  12&   2.2&   1.1&   8.9&   1.5&   1.0&   0.4&   2.311 & 0.226   \\
  11&  Berkeley 4        & 122.28&   1.53&  3200&  0.70& 7.08&  0.08&  0.20& 3&
4&  19&   4.3&   2.0&  14.6&   2.7&   0.5&   0.2&     2.815 & 0.244 \\
 13&  Alessi 1          & 123.26& $-$13.30&   800&  0.10& 8.85&  0.13&  0.45&4&
1&  19&   1.0&   0.3&   9.1&   2.3&   3.2&   1.3&   2.390 & 0.340   \\
 15&  Platais 2         & 128.23& $-$30.57&   201&  0.05& 8.54&  1.70&  1.70& 2&
2&  18&   4.4&   1.2&   8.9&   1.1&   1.2&   0.4&   2.411 & 0.166   \\
 17&  NGC 457           & 126.63&  $-$4.37&  2429&  0.47& 7.38&  0.12&  0.26& 4&
2&  26&   1.2&   0.6&  13.1&   3.1&   2.6&   1.6&   2.729 & 0.310   \\
 22&  NGC 663           & 129.46&  $-$0.94&  1952&  0.78& 7.14&  0.10&  0.30& 4&
1&  24&   3.1&   1.1&  15.3&   3.3&   0.8&   0.2&   2.965 & 0.280   \\
 23&  Collinder 463     & 127.28&   9.40&   702&  0.30& 8.35&  0.22&  0.72& 4&
3&  77&   3.2&   0.4&  12.4&   1.1&   3.3&   0.3&     2.803 & 0.124 \\
...&  ...               & ...   & ...     & ...  & ...  & ... & ...  & ...
&..&..&....&  ... & ...  & ...  & ...  & ...  & ...  & ...     & ...     \\
1128& ASCC 128           & 109.93&  $-$5.96&   900&  0.13& 8.44&  0.15&  0.35&
2&1&  11&   6.7&   3.5&   7.8&   0.9&   1.9&   1.3&   2.215 & 0.156  \\
\hline\hline
\end{tabular}
\end{table*}

A relative random error of the cluster mass
$\delta_{M_{c}}=\varepsilon_{M_{c}}/M_{c}$ can be derived
from eq.~(\ref{eq:mcl}) as
\begin{equation}
\delta_{M_{c}}^{2}=9\,\delta_{r_{t}}^{2}+\left(\frac{2A-B}{A-B}\right)^{2}
\delta_{A}^{2}+\left(\frac{B}{A-B}\right)^{2}\delta_{B}^{2} \label{eq:de_mcl}
\end{equation}
where $\delta_{r_{t}}=\varepsilon_{r_{t}}/r_{t}$,
$\delta_{A}=\varepsilon_{A}/A$,
and $\delta_{B}=\varepsilon_{B}/B$ stand for relative random errors of
the tidal radius and of Oort's constants, respectively.
The cluster masses and their relative errors are computed with 
  eqs.~(\ref{eq:mcl}) and~(\ref{eq:de_mcl}), and are discussed in Sec. 4.3.

Eq.~(\ref{eq:de_mcl}) can be also used to get a ``rule of thumb'' for
the prediction of the expected accuracy of cluster masses derived
from tidal radii.
Assuming at first $A\approx A_0$ and $B\approx B_0$, one obtains
$\delta{}_{A}\approx 0.06$, $\delta_{B}\approx 0.08$. 
Taking further into account that typically $\delta_{r_{t}}>0.1$ (see
Sec.~\ref{sec:res}),
one finds that the relative error of mass determination is dominated
strongly by the uncertainties of the tidal radius i.e.,
\[
9\,\delta_{r_{t}}^{2}\gg\left(\frac{2A-B}{A-B}\right)^{2}\delta_{A}^{2}+\left(
\frac{B}{A-B}\right)^{2}\delta_{B}^{2},
\]
and therefore
\begin{equation}
\delta_{M_{c}}\approx3\,\delta_{r_{t}}.\label{eq:dlt_mcl}
\end{equation}

\section{King's parameters: Results}\label{sec:res}

\subsection{Construction of the output sample}\label{sec:construc}

For 236 out of 290 clusters in the input list, we obtained at least one
set of King's parameters by applying the method described in Sec.~3.1. Depending
on the membership group, each of these clusters got from one to four
different solutions, and the total number of solutions was 708.
Per cluster, we have at least one set of parameters which are larger
than their $rms$ errors, i.e. with relative errors of
$\delta_{r_{c}}=\varepsilon_{r_{c}}/r_{c}<1$,
$\delta_{r_{t}}=\varepsilon_{r_{t}}/r_{t}<1$, and
$\delta_{k}=\varepsilon_{k}/k<1$.
Since, for the majority of  clusters,  more than one set of
King's parameters was obtained, we need a decision strategy on the priority
of using the results for further analysis.  From the point of view of
membership, the parameters from 1$\sigma$-members should be the most reliable,
but they are more uncertain from the point of view of statistics due to
the relative low number of stars. The opposite is true for
``4$\sigma$-members''.

In order to compare the solutions derived from different membership groups,
to check possible systematics between them and
to define more or less objective criteria of parameter selection, we
considered a subset of 114 clusters. This subset includes all clusters
having four different solutions.
For each of these clusters, we computed the mean from the four
solutions for
a given parameter and used this mean as a normalizing factor.
If a normalized  parameter
is significantly smaller than one,
we conclude that a given membership group delivers a significantly
smaller parameter than the other groups, and {\it vice versa}.
Table~\ref{tab:solqua} gives
the corresponding normalized parameters averaged over 114 clusters.

The most impressive
feature of Table~\ref{tab:solqua} is that the core radii $\bar{r}_c$
do not depend on the membership group used for computation, and tidal radii
$\bar{r}_t$ show only a slight systematic dependence on the membership groups.
On average, the tidal radii obtained with the ``$4\sigma$''-membership samples
are smaller only by a factor of 1.1 than those with $1\sigma$-members.
On the contrary, the parameter $k$ increases towards the
``$4\sigma$''-membership sample and correlates strongly with the normalized
number of cluster members $n_2$ located within an area of a radius  $r=r_2$.
This is a logical behaviour and follows from the meaning of $k$ and $n_2$ in
eq.~(\ref{eq:kingi}).
A relation between the normalized parameter $k$ and the number
of the sample $i$ ($i = 1,...,4$) used for the solution can be approximated by
\begin{equation}
\bar{k} = (0.18\pm 0.04)\times i + (0.54\pm 0.11). \label{eq:ki}
\end{equation}

As expected, the relative errors in the determination of the parameters
$\bar{\delta}_{r_c}$, $\bar{\delta}_{r_t}$, and $\bar{\delta}_{k}$
are largest for the $1\sigma$-solution. The $1\sigma$-sample
contains the most probable cluster members, but their number is relatively low
compared to the other membership samples. This is the reason for
relatively large Poisson errors and, consequently, for higher $rms$
errors in the fitted parameters.

The goodness of a fit is given by the $Q(\chi^2,\nu)$-probability,
an output parameter of the fitting pipeline (see Sec.3.1).
The normalized $\overline{Q}(\chi^2,\nu)$-parameters averaged over 114 clusters
are also given in Table~\ref{tab:solqua}. As can be seen,
$\overline{Q}(\chi^2,\nu)$ does not depend on the richness of the sample and
indicates a more suitable fitting with $1\sigma$- and $2\sigma$-samples
than with ``$4\sigma$''-groups.

Based on the statistics in Table~\ref{tab:solqua}, we chose the
following ranking of the solutions. We give the highest weight to the
solutions with the largest $Q(\chi^2,\nu)$-probability.
If, for a given cluster there are more than one solutions of the same
quality (i.e., the $Q(\chi^2,\nu)$-probabilities differ by less than
0.1\%), we use an additional criterion based on $\chi^2$-values
supplied by the fitting pipeline. Since $\chi^2$ does depend on the sample size
$n$ and on the degree of freedom $\nu$ , a readjusting of
the $\chi^2$-estimate is needed
when we compare fitting results derived with different membership samples.
Therefore, we select a solution with a smaller
value $\chi^2_{\nu,n} = \chi^2/(\nu \times n)$ which
is, in fact, an average mean square deviation of the observed from the fitted
profiles in
units of Poisson errors computed per star.
In a few cases when even the $\chi^2_{\nu,n}$-parameters differ insignificantly
(by less than 0.001) for two solutions, we
give priority to the solution with smaller error $\varepsilon_{r_t}$.

According to the selection procedure, the solution from the $1\sigma$-samples
gets the best ranking in 94 cases out of 236,
from $2\sigma$-samples - in 72 cases,
from $3\sigma$-samples - in 39 cases,
and ``$4\sigma$''-samples -
in 31 cases.

The data on the structural parameters are compiled in a table which is
available in
machine-readable form only. As an example,
Table~\ref{tab:results} lists a few entries of the complete data.
Column 1 gives the cluster number in the COCD catalogue,
columns 2 through 9 are taken from the COCD, while columns 10 through
20
include the information obtained in this work. For each of the 236 clusters we
give:
name (2), galactic coordinates (3, 4), the distance from the Sun in pc
(5), the
reddening (6), the logarithm of the cluster age in years (7), empirical angular
radii (in
degrees) of the core $r_1$ (8) and of the corona $r_2$ (9). Column (10) is the
number of
the acceptable solutions of King's parameters from the four membership groups.
Column (11) gives  the number of the membership sample ($1\sigma$, $2\sigma$,
$3\sigma$,
or ``$4\sigma$'') chosen by the selection procedure as providing the best
solution,
whereas column (12) is the number of cluster members of this sample within $r_2$
after
background correction. Columns 13 through 18 give the corresponding King
parameters
$r_c,\,r_t,\,k$ with their $rms$ errors. The parameters are taken
as  obtained from the selected solution. Depending on the applications desired,
the reader is
advised to take into account the empirical relations between the selected
solution for $r_t$ and $k$ and  the membership sample used to obtain these
parameters
(cf. Table~\ref{tab:solqua} and eq.~(\ref{eq:ki})).
Finally, columns (19, 20) provide the logarithm of the cluster mass and its
$rms$
error computed
from $r_t$ by use of eq.~(\ref{eq:mcl}) and discussed in Sec.\ref{sec:mass}.

\subsection{Properties of the sample of King's parameters for open clusters}

\begin{figure}
\includegraphics[width=1.0\columnwidth]{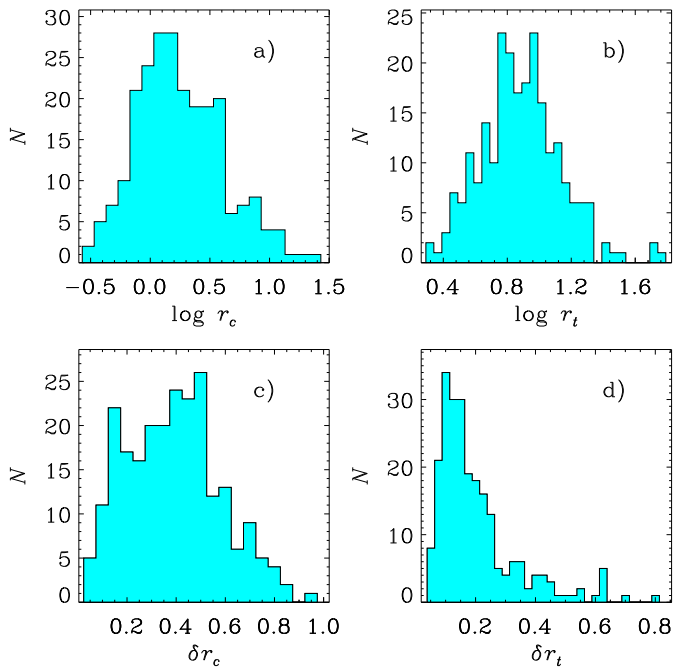}

\caption{Distributions of King's radii and their relative errors for
236 open clusters. Panels \textbf{(a)} and \textbf{(b)} are for
the core radius $r_c$ and the tidal radius $r_t$ measured in pc.
Distributions of relative errors in core radius $\delta_{r_c}$ and
in tidal radius $\delta_{r_t}$ are shown in panels \textbf{(c)} and
\textbf{(d)}, respectively.}
\label{fig:parerd}
\end{figure}

\begin{figure*}
\includegraphics[%
  clip,
  width=1.0\textwidth,
  keepaspectratio]{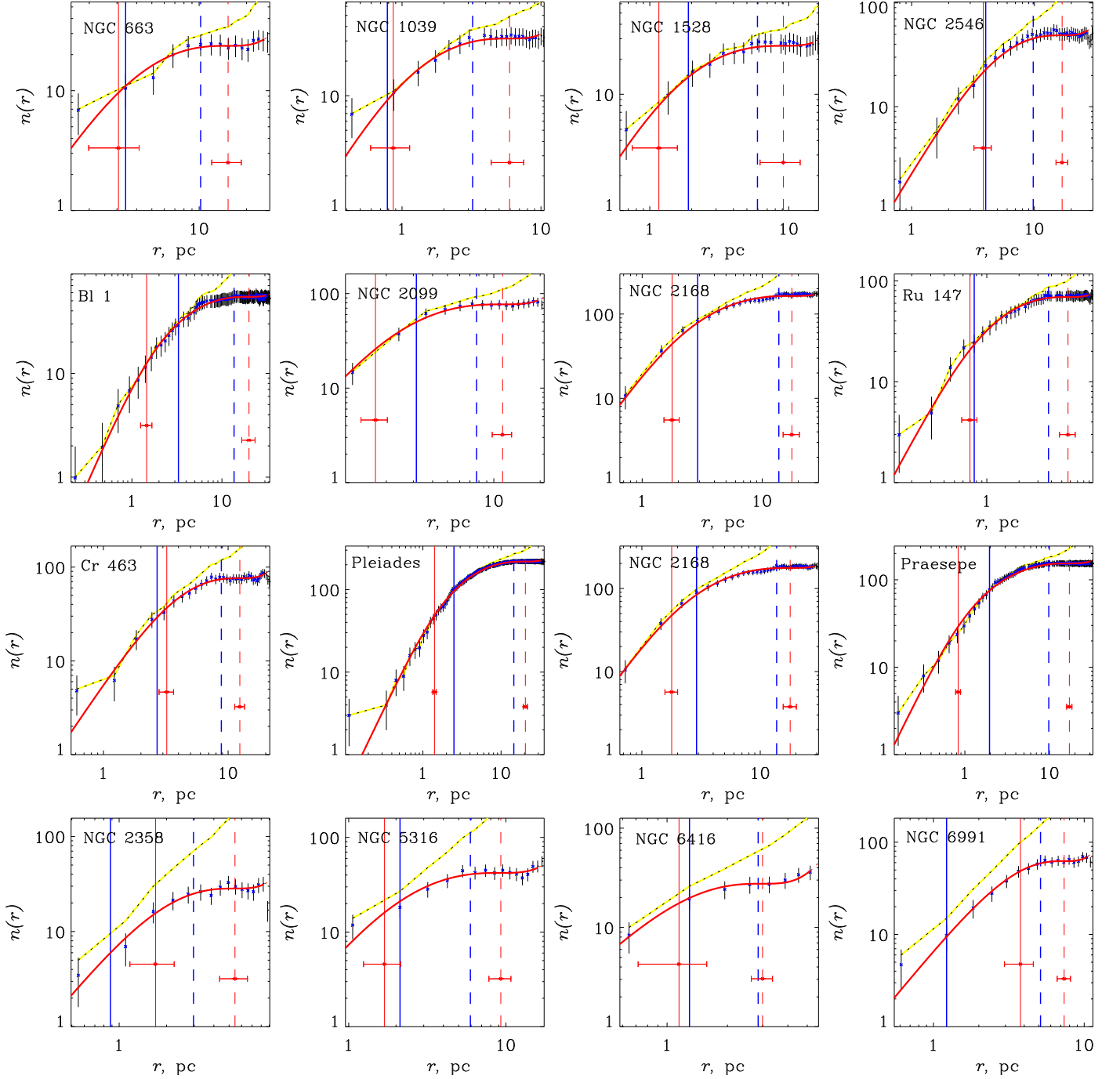}

\caption{Examples of radial profiles from our final sample ordered by the
priority
of the solution. The uppermost row represents the $1\sigma$-sample.
The second row is for
$2\sigma$-solutions, the third and forth rows illustrate $3\sigma$ and $4\sigma$
samples. The crosses show empirical data corrected for the background, the bars
are Poisson errors. The vertical lines indicate the derived radii. The solid
lines show $r_2$ and $r_c$, the broken lines $r_2$ and $r_t$. The empirical
parameters are shown with blue color, while the fitted data are shown with red
color. Horizontal bars indicate the value of the $rms$ error of the parameter.
The solid curve is the fitted King profile, the broken curve (yellow) is the
total profile not corrected for background.}\label{fig:multi}
\end{figure*}

In Fig.~\ref{fig:parerd}(a,b) we show the distribution of 236 clusters over
the derived parameters $r_c$ and $r_t$. The distributions are rather compact
with peaks at about 1.5 pc for the core radii and at about 7 - 10 pc
for the tidal radii. According to Fig.~\ref{fig:parerd}(c,d),
the relative error $\delta_{r_c}$ of the core radius is typically
between 15\% and 50\%,
whereas the tidal radius is more accurate: the majority of clusters has
the tidal radius determined with a relative error $\delta_{r_t}$ better than
20\%.
Therefore, for these clusters, we expect to obtain reasonable estimates of
masses
from the present data (cf. eqs.~(\ref{eq:de_mcl}), (\ref{eq:dlt_mcl})).
As a rule, clusters which are more distant and/or less populated got
relatively large errors in $r_t$. In these cases the method reaches its
limitations. Probably, if applied to deeper photometric data, the
method may provide acceptable results for a large portion of these
clusters. Nevertheless, there is a number of clusters with ``irregular''
density profiles, and, fitting them by a model  proposed for spherical systems
in
equilibrium, does not have  much prospect of success.

In order to give the reader an idea of typical profiles we show, in
Fig.~\ref{fig:multi}, a set of different cases selected from
the final sample. Compared to the empirical cluster radius $r_2$,
the tidal radius $r_t$ ranges between $1\,r_2$ and $2\,r_2$.
For clusters with relatively accurate radii ($\delta_{r_c}< 0.33$
and $\delta_{r_t} <0.33$), the averaged relation is
\[
r_t = (1.54\pm0.02)\times r_2.
\]
Therefore, we conclude that the empirical parameter $r_2$ scales
the tidal radius reasonably well.

\subsection{Cluster masses}\label{sec:mass}

\begin{figure}
\includegraphics[width=1.0\columnwidth]{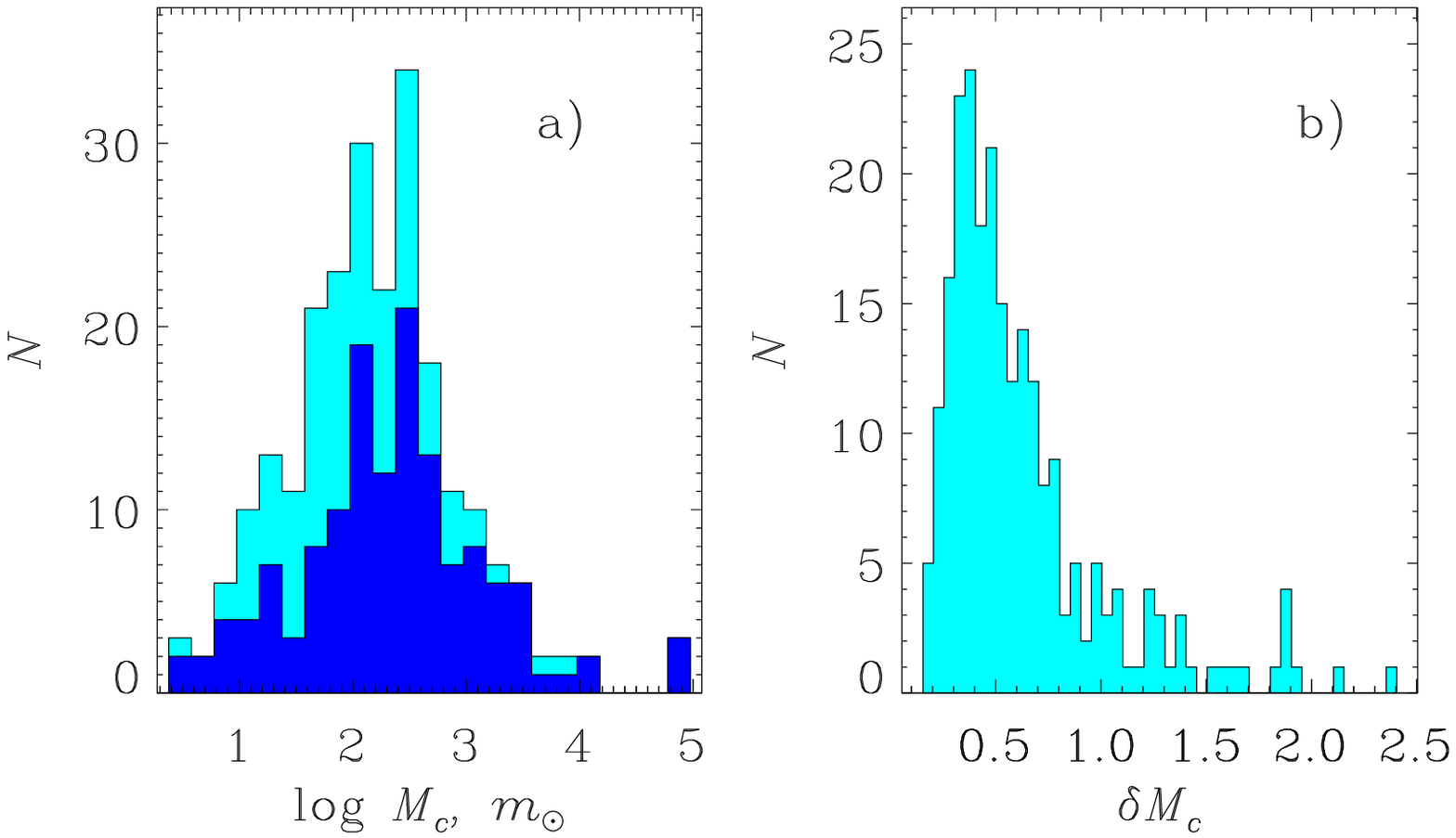}

\caption{Distributions of cluster masses (panel \textbf{(a)}) and their relative
$rms$ errors(panel \textbf{(b)}). Clusters with accurate masses($\delta_M<
60\%$) are shown as the dark histogram in panel (a).
}\label{fig:mass}
\end{figure}

Using eq.~(\ref{eq:mcl}) and the tidal radius $r_t$, we estimated
masses for each of the 236 open clusters. The results are shown in
Fig.~\ref{fig:mass}
where the distribution of clusters over mass is given in panel a), whereas
panel b) shows the distribution of relative errors in mass. Most of the
clusters in our sample have masses in a range $\log M_{c}/m_\odot= 1.6 - 2.8$,
though a few clusters have  masses as small as $10\,m_\odot$. Three objects got
masses of about  $10^5\,m_\odot$. They are the associations
Nor~OB5, Sco~OB4 and Sco~OB5. For 139 clusters, the masses were obtained with
a relative accuracy better than 60\%. Their distribution shows the same
features as the complete sample.

We note that the masses based on $r_t$ from eq.~(\ref{eq:kingi}) do
not take into account a possible flattening of clusters which arises due to the
tidal
coupling with the
Milky Way. 
Because of this, a stellar cluster has a shape of a three-axial
ellipsoid  with
the major axis oriented  in the direction of the Galactic
centre. In general, we 
obtain 
a projection of the tidal radius on the celestial
sphere 
from eq.~(\ref{eq:kingi}), 
and the
relation between the tidal radius and this projection  depends on the mutual
position of the Sun and a given cluster. 
Comparing masses of two clusters 
with different location in the Galactic disk, the corresponding effect must,
therefore, be taken into account.

\section{Comparison with other determinations}\label{sec:compar}

\subsection{King's radii from a three parameter fit}

In contrast to massive spherical systems, there are only a few results reported
on the determination of
structural parameters of open clusters via direct parameter fits of King's
profiles to the observed density distributions. Some of them consider
remote clusters (e.g., King~\cite{king62}, Leonard~\cite{leo88}), or
newly detected cluster candidates (Froebrich et al.~\cite{froeb07})
which are absent
in our cluster sample. Others are based on a two parameter fit
(Keenan~\cite{keen73},  Bica~\cite{bibob}, Bonatto et al.~\cite{bonb}), and,
therefore, these results cannot be compared directly with ours.
We found only seven papers where a three parameter fit was applied to
open clusters. In these papers only four clusters are in common with
our sample. These are
two nearby clusters, the Pleiades and Praesepe, NGC~2168 (M35) at 830~pc
from the Sun, and the relatively distant cluster NGC~2477, at 1.2~kpc.

\begin{table*}
\begin{minipage}[t]{\textwidth}
\caption{Comparison of our results with literature data on King's parameters for
the clusters in common}\label{tab:compar}
\begin{tabular}{lrrrrrrccccc}
\hline\hline
Cluster &
\mc{5}{c}{Present paper}&
\mc{6}{c}{Literature}\\
 &$n_2$&
  $r_{fit}$&
 \mc{1}{c}{$r_c$}  &
 \mc{1}{c}{$r_t$}  &
 \mc{1}{c}{$M_{c}$} &
 $N^*$&$r_{area}$&
 $r_c$  &
 $r_t$  &
 note &
 ref\\
        &
        &
\mc{1}{c}{deg}&
\mc{2}{c}{pc}&
 \mc{1}{c}{$m_\odot$}&
        &
  deg   &
\mc{2}{c}{pc} &
  &
 \\
\hline
%
Pleiades&219&14   &$1.4\pm0.1$&$20.5\pm1.3$&$3107\pm638$&    270&  6  &
$1.4\pm0.5$        & $16\pm7$&     &1\\
        &   &     &          &             &            &   1067&3    &0.9...2.9
  & 13.1&$a$  &2\\
        &   &     &          &             &            &   1200& 10  &2.1...2.8
  &12.4...14.5&$b$ &3\\
Praesepe&154& 9   &$0.8\pm0.1$&$17.1\pm1.2$&$1806\pm428$&    185&  4  &
$1.0\pm0.5$& $11.1\pm4.9$&     &4\\
        &   &     &         &             &            &  ~1000&3.8  &3.5
 &16&$c$    &5\\
NGC~2168&186& 1.95&$1.8\pm0.2$&$17.9\pm2.3$&$724\pm677$ &       &0.47 &
1.3...4.3  &$8.35...\infty$ &     &6\\
NGC~2477& 10& 0.55&$1.9\pm1.0$&$6.5\pm1.2$ &$89\pm51$   &       &0.25 & 1.8
  &  8.1&$d$  &7\\

\hline
\end{tabular}\\
\textit{Notes: }$(a)$ -- $r_c$ is computed for member groups of different mass,
$r_t$ is
derived from the cluster mass via eq.~(\ref{eq:mcl}); $(b)$ -- $r_c,\,r_t$ are
obtained from
the distribution of low-mass stars ($m<1\,m_\odot$); the original results  are
given
in angular units and are transformed to linear sizes by us;
$(c)$ -- $r_c,\,r_t$ are obtained from the distribution of low-mass stars
($m<1\,m_\odot$); $(d)$ -- the area is a square of $0.5\times0.5$~sq.deg.\\
\textit{References: }1-- Raboud \& Mermilliod~(\cite{ramer98a}), 2-- Pinfield et
al.~(\cite{pin98}), 3-- Adams et al.~(\cite{adea01}), 4-- Raboud \&
Mermilliod~(\cite{ramer98b}), 5-- Adams et al.~(\cite{adea02}), 6--
Leonard \& Merritt~(\cite{leo89}), 7-- Eigenbrod et al.~(\cite{eigen}).
\end{minipage}
\end{table*}

In Table~\ref{tab:compar} we compare the results obtained in this paper
with those found in the literature for the four clusters. Column 1 is the
cluster identification. In columns 2 through 6 we provide our results:
in column 2 we give the number of members within the empirical cluster radius
$r_2$, column 3 is for $r_{fit}$ i.e. a radius found as best suited
for the profile fitting (cf. Sec. 3.1). The results of the fitting,
King's radii $r_c$ and $r_t$ are listed in columns 4 and 5  together
with their $rms$ errors, whereas the corresponding tidal mass is given in
column 6. Columns 7 through 10 show the data from literature: columns 7
and 8 give the number of stars used to construct the density profiles and
the radius of the area considered, respectively. The King radii are listed in
columns 9 and 10. Note that  we are
not able to keep a uniform format for these data due to the different
presentation of the results in different papers.

We found three different estimates of the tidal radius published for the
Pleiades. A direct comparison with our results can be done
only with radii obtained by Raboud and Mermilliod~(\cite{ramer98a}) who
consistently applied
the 3-parameter fitting technique both to various sub-samples of
the Pleiades stellar population and to the total membership sample. From 
Table ~\ref{tab:compar} we
conclude that their findings of $r_c$ and $r_t$ coincide well with our results.
A slightly different method was applied by
Pinfield et al.~(\cite{pin98}) to derive a tidal radius for the Pleiades.
They used differential density profiles, which were constructed
for cluster members falling
in different mass ranges, and tidal radii computed for each of these
sub-samples.
The results were used to derive the partial and total masses of the cluster.
The tidal radius for the cluster as a whole was then computed from the cluster
mass
via eq.~(\ref{eq:mcl}).
A similar approach was used by Adams et al.~(\cite{adea01}) who applied
a 3-parameter fit to a sample of low mass ($m<1\,m_\odot$) members of the
Pleiades. Again, the tidal radius $r_t$ was derived from stellar mass counts in
the cluster by eq.~(\ref{eq:mcl}). The approach by Pinfield et
al.~(\cite{pin98})
and Adams et al.~(\cite{adea01}) provides a smaller value for $r_t$ and  larger
value for $r_c$ (for Pleiades members less massive than $1.2\,m_\odot$)  than
a direct 3-parameter fitting of the density distribution applied to the full
sample of cluster members. Nevertheless, taking into account that
the three studies and ours are based on observations of different spatial and
magnitude coverage, and on an independent membership evaluation, the
agreement between the results is quite acceptable.

We arrive at similar conclusions in the case of Pra\-e\-se\-pe.
Our estimate for $r_t$ is compatible with the result by
Raboud \& Mermilliod~(\cite{ramer98b}) within the $rms$ errors, it
coincides well with the finding of Adams et al.~(\cite{adea02}) who, as in
the case of the Pleiades,
fitted the King profile to low-mass ($m=0.1-1\,m_\odot$) cluster members.
With respect to $r_c$, we achieved good agreement with the estimate by
Raboud \& Mermilliod~(\cite{ramer98b}) whereas $r_c$ from
Adams et al.~(\cite{adea02}) is significantly higher.
Such a systematic difference can possibly be explained by the considerably
deeper
survey used by Adams et al.~(\cite{adea02}) in studies of the Pleiades and the
Praesepe. Since their empirical profiles are based on USNO POSS~I E
and POSS~II F plate scans, and on the 2MASS catalogue, the input data
are dominated by lower mass stars which, due to the energy equipartition
process,
show a wider distribution than the more massive stars.

For the last two clusters in Table~\ref{tab:compar}, there are only two papers
reporting the determinations of King's parameters. Leonard \&
Merritt~(\cite{leo89})
studied the central area of the NGC~2168 cluster. Therefore, they were able
only to set probable limits for the cluster radii. Our estimates of
$r_c$ and $r_t$ fit these limits well. For the relatively distant and rich
cluster NGC~2477, the King radii were published by Eigenbrod et
al.~(\cite{eigen}).
The authors defined the membership sample on the basis of radial velocities
of numerous red giants and constructed density profiles for groups of stars
of various masses. The core and tidal radii of the cluster were determined from
the comparison of the structure parameters of single groups. Also in this case,
their results are in good agreement with our estimates of $r_c$ and $r_t$,
though
NGC~2477 belongs to the poorer clusters in our sample due to its large distance
from
the Sun.
With 10 members within $r_2$, NGC~2477 satisfies marginally our
constraints. Nevertheless, the fitted parameters coincide well with
the corresponding estimates obtained with the much deeper survey ($V<17$)
by Eigenbrod et al.~(\cite{eigen}).

\subsection{Various scales of cluster masses}

Although the adopted values of Oort's constants and of a distance to a cluster
can slightly influence its mass estimate, the tidal radius is the major source
of
uncertainty in the mass determination via eq.~(\ref{eq:mcl}), simply
due to the cubic power relation between tidal radius and cluster mass.
Analysing different groups of the Pleiades members, Raboud
\& Mermilliod~(\cite{ramer98a}) concluded that the tidal mass of the
Pleiades is about $1400 m_\odot$ with a $1\sigma$ confidence interval of
$[530,\,2900]\,m_\odot$. With Oort's constants $A = 15$~km/s/kpc, $B = -12$
km/s/kpc
and a cluster distance of 125~pc adopted by the authors, a strict application
of eq.~(\ref{eq:mcl}) would provide a tidal mass of $1620 m_\odot$
$([330,\,4600]\,m_\odot)$
for the Pleiades. From a similar analysis for Praesepe,
Raboud \& Mermilliod~(\cite{ramer98b}) derived a tidal mass of $440\, m_\odot$
$([157,\,987]\,m_\odot)$ (a consequent use of eq.~(\ref{eq:mcl})
would give $520 m_\odot$ $([90,\,1570]\,m_\odot)$).
We conclude that the disagreement between our estimates and those of
Raboud \& Mermilliod~(\cite{ramer98a},
\cite{ramer98b}) for the Pleiades and Praesepe
are mainly caused by statistical uncertainties in the determination
of the tidal radii (cf. Table~\ref{tab:compar}).

In the case of NGC~2477, the situation is less clear.
For this cluster, Eigenbrod et
al.~(\cite{eigen}) estimated the tidal and virial masses to be
$M_{tid}=5400\,m_\odot$ and $M_{vir}=5300\,m_\odot$, respectively.
This result differs considerably from our estimate.
Moreover, it is in contradiction to the value of their tidal radius of
$r_t=8.1$~pc
which according to eq.~(\ref{eq:mcl}), should give a tidal mass of
 $M_{tid} \approx 170\,m_\odot$.
Therefore, the coincidence achieved by Eigenbrod et al.~(\cite{eigen})
between the tidal and virial masses should be considered with caution.
We note that
NGC~2477 is still badly studied, and its
membership is poorly established. Therefore, a contamination of the stellar
sample from Eigenbrod et al.~(\cite{eigen}) is well probable. This could be one
reason for an overestimation of the velocity dispersion and, consequently,
of the virial mass of the cluster. The presently large uncertainties in
kinematical data
for determing virial masses can also cause these discrepancies.
(see Appendix~\ref{sec:mvir} for more detail).
On the other hand, this large disagreement can be partly explained by an
underestimation of the tidal radius:
if this relatively distant cluster is  subject to  strong mass segregation,
low mass stars on the cluster edges can be beyond the magnitude limit even in a
deep survey.

\begin{figure}
\includegraphics[width=0.9\columnwidth,bbllx=40,bblly=84,bburx=355,bbury=685,
clip=
]{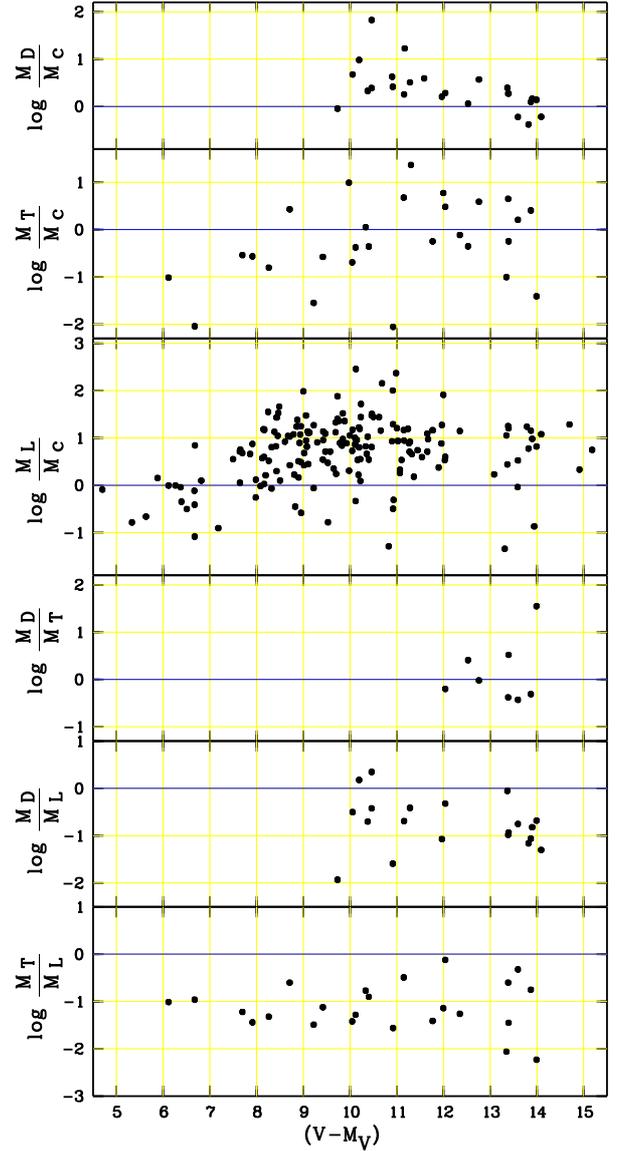}
\caption{Comparison of tidal masses $M_c$ of open clusters derived in the
present
study with literature values on counted- and MF-scaled masses as
a function of the
distance modulus. From top to bottom we show ratios of: counted masses
of Danilov \& Seleznev~(\cite{danil}) ($M_D$ to tidal masses); counted masses
of Tadross et al.~(\cite{tad02}) ($M_T$ to tidal masses); and the IMF-scaled
masses of Lamers et al.~(\cite{lamea}) ($M_L$ to tidal masses). The last
panels show intercomparisons of literature data.}\label{fig:mascomp}
\end{figure}

Since literature data on tidal masses of open clusters are rather scarce, we
looked for recent publications on cluster masses estimated with other methods.
We omit here a discussion on the determination of virial masses of open clusters
and refer the reader to Appendix~\ref{sec:mvir} where this method is
discussed in more detail. In order to compare our results on cluster masses,
we consider only those publications where cluster masses are obtained for
a relevant number of open clusters rather than for a single cluster.
Under these constraints, we found three publications on mass determination
for galactic open clusters, Danilov \& Seleznev~(\cite{danil}),
Tadross et al.~(\cite{tad02}), and Lamers et al.~(\cite{lamea}).
In Fig.\ref{fig:mascomp} we compare the different results on mass estimates.

Danilov \& Seleznev~(\cite{danil}) derived masses for 103 compact, distant
(>1 kpc) clusters from star counts down to $B=16$ from homogeneous
wide-field observations with the 50-cm Schmidt camera of the Ural university.
For each cluster, the authors estimated the average mass of a star
observed in a cluster and then computed the total visible cluster mass.
The average mass is found either from star counts, or from an
extrapolation of the Salpeter IMF down to the magnitude limit $B=16$.
On one hand, their cluster masses should be underestimated due to
their magnitude limited survey, and the bias
should increase with increasing distance modulus of a cluster. On the other
hand,
without membership information, the masses could be overestimated for clusters
located at relatively low distances from the Sun. In a certain respect,
these biases may partly compensate each other.

Based on UBV-CCD observations compiled from the literature, Tadross
et al.~(\cite{tad02}) redetermined ages and distances for 160 open clusters, and
derived cluster masses from counts of photometrically selected cluster members.
Since they used observations taken with different telescopes,
i.e., for different clusters one expects different limiting magnitudes, it is
rather
difficult to estimate possible biases. In any case, a large portion of clusters
should get underestimated masses due to the relatively small area of the sky
usually covered by CCD observations in the past.

Lamers et al.~(\cite{lamea}) used data from the COCD for the determination
of cluster masses. For each cluster, the authors normalized the Salpeter IMF
in the mass range of stars present in COCD ($V < 11.5$), and the normalized
Salpeter IMF
was extrapolated from large masses down to  $m=0.15\,m_\odot$.
For distant clusters with $V-M_V>8$, the extrapolation was done over a
very large range, from masses larger than $\approx 1.5\,m_\odot$ (or $M_V<3.5$)
to masses of $0.15\,m_\odot$ (or $M_V \approx 13$), where the IMF is
still not very well known.
If the IMF at low masses were flatter than the Salpeter IMF
(see e.g. Kroupa et al.~\cite{krea93}),
the approach by Lamers et al.~(\cite{lamea}) would give overestimated
cluster masses. Moreover, in a segregated cluster one should expect different
forms
of the mass function in the central area and at the edges. Integration
of the Salpeter IMF over the complete cluster area would, also, result in
overestimating the cluster mass. A comparison of our cluster masses with
those of Lamers et al.~(\cite{lamea}) is especially interesting: since
both papers use the same observational basis, we can estimate uncertainties
caused by the different approaches.

According to our determination of cluster masses based on
tidal radii, we expect possible biases for  relatively
distant clusters where we observe only a tip of bright stars.
Especially in segregated clusters, due to energy equipartition
the brightest stars are more concentrated to the cluster centre,
and do not reproduce the correct tidal radius.
Nevertheless, for distant clusters we can see a distance-dependent
effect only in the upper panel of Fig.\ref{fig:mascomp}, where we
compare our masses with those of Danilov and Seleznev~(\cite{danil}).
Since the $M_D/M_C$-relations decrease with distance
modulus, we suppose that our method manages biases better
than the approach by Danilov and Seleznev~(\cite{danil}).
Moreover, the dependence of the $M_D/M_C$-relations on distance modulus may be
explained alone by the biases in the mass determination by
Danilov and Seleznev~(\cite{danil})
described above, i.e. overestimated masses for clusters at low distances and
underestimated masses for distant clusters. This interpretation seems to
be plausible,
but we cannot exclude completely distance-dependent biases in
our determinations.

Furthermore, due to a possibly elongated form of open clusters
(cf. Sec.~\ref{sec:detrm}),
our estimates provide a lower limit for cluster masses.
This can be one of the reasons for systematic differences to cluster masses from
Lamers et al.~(\cite{lamea}). Except for nearby clusters with $V-M_V<8$,
their masses are, on average, larger by a factor of 10. On the
other hand, masses from Lamers et al.~(\cite{lamea}) are systematically
larger than masses from Danilov and Seleznev~(\cite{danil}) and
Tadross et al.~(\cite{tad02}), too. Therefore, we can not exclude that the
approach by Lamers et al.~(\cite{lamea}) is, at least partly, responsible
for these differences.

\section{Conclusions}\label{sec:concl}

In an ideal case, i.e. if, for a given cluster, the membership is determined
with certainty and completeness,
the cluster mass could simply be derived from counting masses of
individual members. In the future, with deeper surveys and an increasing
accuracy of kinematic and photometric data, this primary method will
provide sufficiently accurate and uniform mass estimates for a significant
number of
open clusters in the Galaxy. At present, however, there is no way to
measure the masses of open clusters directly. The methods currently applied
require a
number of assumptions, and depending on the method and assumptions used, the
results
can differ by a factor of 100 for individual clusters (cf.
Fig.\ref{fig:mascomp}).
Therefore, the determination of cluster masses is still
a very challenging task.

Our aim was to estimate masses for a larger number of clusters by applying
a uniform and possibly objective method, and to obtain an independent
basis for statistical studies of the distribution of cluster masses in the
Galaxy.
In our work we could benefit from the homogeneous set of cluster parameters
derived for 650 open clusters with good membership based on the astrometric and
photometric data
of the \ascc. The estimation of cluster masses was done via tidal radii
determined from a three-parameter fit of King's profiles to the observed density
distribution (King~\cite{king62}).
This method is weakly dependent on assumptions and can be applied to all
spherical
systems in equilibrium with well-defined density profiles. Since these
requirements
are not always met in the case of open clusters, we could obtain solutions
only for 236 clusters, i.e. for less than half of the clusters in our sample.
However, this
number is considerably larger compared to the small number of clusters with
tidal masses determined before.

The main difficulties in the practical application of King's model
to open clusters arise from
the relatively poor stellar population (compared to globular clusters) and
from the higher degree of contamination by field stars in the Galactic disk.
Using an all-sky survey, we could rely  on the completeness of data in
the selected sampling areas, down to the limiting magnitude of the \ascc.
Further, since we were free in selecting the size of the sampling areas,
we were able to optimize the boundary condition for each cluster as
well as possible. As a result, we could partly decrease the influence of the
above mentioned problems in applying King's method and improve the solutions
by taking into account the outermost regions of the clusters and by
excluding residual field stars from the solution.

Together with a realistic membership based on both kinematic and
photometric constraints, a good profile fitting could be achieved even for
clusters with a relatively low number of members.
The highest quality of the fitting (goodness-of-fit) was achieved
with the best-determined membership sample
(so-called ``$1\sigma$''-members) and, hence, a low contamination by field
stars.
However, it turned out that the membership criterium alone did
not have very strong impact onto the values of the
fitted parameters
$r_c$ and $r_t$ themselves. In fact, for compact and
relatively distant clusters,
we sometimes found the best results without a
preliminary membership selection (so called ``$4\sigma$''-members).
In conclusion, this paper could be seen as a justification for a simple
application
of King's method to observed brightness profiles of compact open clusters no
matter if
membership is determined or not, provided that the observed density profiles
are properly corrected for the background.

\begin{acknowledgements}
We are grateful to Henny Lamers for providing us with unpublished
data on cluster masses. This study was supported by DFG grant 436 RUS 113
/757/0-2, and RFBR grant 06-02-16379.
\end{acknowledgements}

\begin{appendix}
\section{Virial masses of open clusters: current status}\label{sec:mvir}

\begin{figure}
\includegraphics[width=1.0\columnwidth,bbllx=80,bblly=104,bburx=510,bbury=610,
clip=]{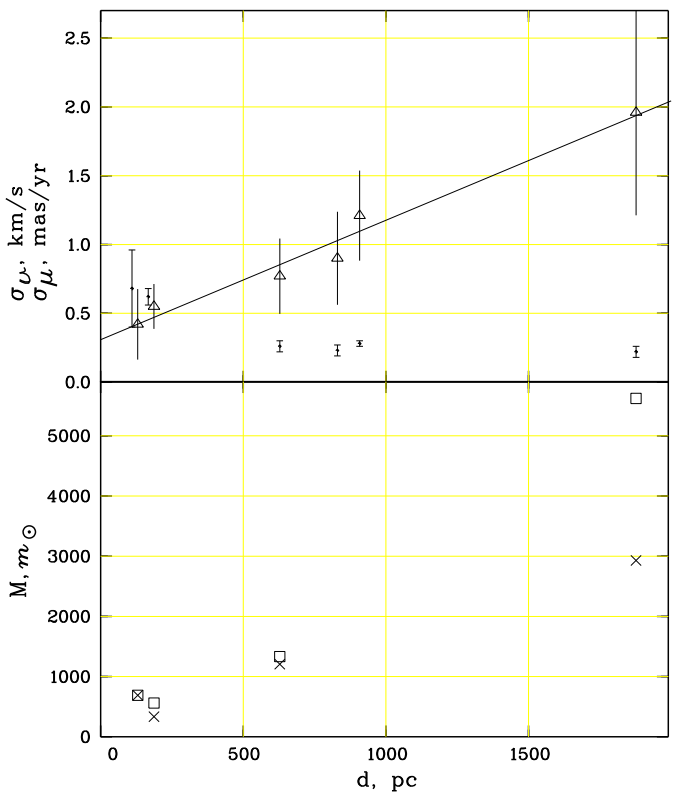}

\caption{Masses and internal velocity dispersions
for the open clusters: Pleiades, Praesepe, NGC~6494, NGC~2168, NGC~2682,
and NGC~6705 (the clusters are ordered with distance).
Upper panel: internal velocity dispersion versus distance. Dots are for proper
motion dispersions, triangles for tangential velocity dispersions.
Bars are $rms$-errors of the
dispersions: bars with hats are for proper motions and bars without hats
for tangential velocities (for the Pleiades and Praesepe, the symbols for
the proper motion dispersions are shifted to the left for a better visibility).
The line shows a linear fit to the tangential velocity dispersion data.
Lower panel: Cluster mass versus distance (based on literature data, see text
for more detail). Crosses mark virial masses obtained from
proper motion dispersions, and squares show counted masses.}\label{vm:fig}
\end{figure}

All our attempts failed to compute reasonable cluster masses from the dispersion
of proper motions and/or radial velocities taken from the \ascc catalogue for
cluster members.
The main reason is the accuracy of kinematical data, which is still
too low in current all-sky surveys.

Up to now, the best published data on velocity dispersions
were obtained for a few open clusters from proper motions obtained from
long-term observations with
the Yerkes 40-inch refractor (F = 19.3 m, a scale of 10.7 arcsec/mm,
a typical epoch difference of more than 55 years, and a typical $rms$-error
of the proper motions of a few 0.1 mas/y).
The clusters are the Pleiades
(Jones~\cite{jones70}), Praesepe (Jones~\cite{jones71}), NGC~6705
(McNamara \& Sanders~\cite{mcn77}), NGC~6494 (McNamara \&
Sanders~\cite{mcn83}), NGC~2168 (McNamara \& Sekiguchi~\cite{mcn86}),
NGC~2682 (Girard et al.~\cite{gir89}).
In these papers, the internal proper motion dispersions are corrected
for different biases, and virial as well as counted masses are determined
for four clusters i.e., for the Pleiades, Praesepe, NGC~6494, NGC~6705.
The results are shown in Fig~\ref{vm:fig}.

Independent of the methods of mass determination, the masses in
Fig.~\ref{vm:fig} show a correlation with cluster distance. Since
the number of clusters is very low, the correlation does not need to be
real, but can appear by chance, due to the small sample.
In order to understand the effect, we transformed the
proper motion dispersions $\sigma_{\mu}$ published for 6 clusters to
one-dimensional tangential velocity dispersions $\sigma_{v}$ via
$\sigma_{v} = 4.74\, d\, \sigma_{\mu} $, where $d$ is the
distance of a cluster adopted in the original papers.
In the upper panel
of Fig.~\ref{vm:fig} we show $\sigma_{\mu}$ and $\sigma_{v}$ as
functions of distance $d$.
Whereas $\sigma_{\mu}$ is independent from the cluster distance,
the tangential velocity dispersion $\sigma_{v}$ indicates a strong
correlation which is well described by a first-order polynomial.
Therefore, we suppose a non negligible random component in
$\sigma_{v}$ resulting rather from residual $rms$ errors in proper motions
than from the internal velocity dispersion. This is not surprising,
because, for all clusters at distances larger than the Pleiades and Praesepe,
the dispersions are of the order of the $rms$ errors of the proper
motion measurements.
Assuming that the internal
velocity dispersion $\bar{\sigma}_{v}$ were similar for all these clusters,
we obtained $\bar{\sigma}_{v} = 0.31$ km/s by extrapolation of
the linear regression polynomial to $d=0$. In other words, we need
tangential velocities determined with an accuracy better than 0.3~km/s
(or proper motions with an accuracy better than 0.06-0.07 mas/y for
a cluster at $d$ = 1~kpc)
for a more or less reliable estimate of its virial mass. Even the
Hipparcos proper motions with typical $rms$ errors of 1 mas/yr
do not meet this requirement.

With respect to radial velocities, one may suppose that these give
a better basis for the determination of the internal velocity dispersion
since their accuracy does not depend on distance. Recently, radial
velocities have been measured in several open clusters for a sufficient
number of cluster members (see e.g., Eigenbrod et al.~\cite{eigen} for NGC~2477,
or F{\"u}r{\'e}sz et al.~\cite{fur06} for NGC~2264). The derived velocity
dispersions are, however, somewhat too large, 0.93 km/s in NGC~2477, and
3.5 km/s in NGC~2264. This may be a consequence of biases which
still affect the data  and which are very difficult to take into account.
Among them may be the contamination by field stars and by unresolved binaries,
or
motions within stellar atmospheres.

\end{appendix}

\end{document}